\documentclass[num-refs]{wiley-article}
\usepackage[final]{changes}

\definechangesauthor[color=red]{Revision2}

\papertype{Original Article}
\paperfield{Journal Section}
 
\title{Real-time fetAl brain and placental T2* mapping at 0.55T low-field MRI (RAT)}

\author[1,2]{Jordina Aviles Verdera}
\author[1,2]{Sara Neves Silva}
\author[1,2,4]{Raphael Tomi-Tricot}
\author[3]{Megan Hall}
\author[3]{Lisa Story}
\author[1,2]{Shaihan J Malik}
\author[1,2]{Joseph V Hajnal}
\author[1,2]{Mary A. Rutherford}
\author[1,2,5]{Jana Hutter}

\affil[1]{Early Life Imaging Department, School of Biomedical Engineering and Imaging Sciences, King’s College London, London, UK}
\affil[2]{Imaging Physics and Engineering Department, School of Biomedical Engineering and Imaging Sciences, King’s College London, London, UK}
\affil[3]{Women's Health, GSTT, London, UK}
\affil[4]{MR Research Collaborations, Siemens Healthcare Limited, Camberley, UK}
\affil[5]{Smart Imaging Lab, Radiological Institute, University Hospital Erlangen, Erlangen, Germany}
\corraddress{Jordina Aviles Verdera, King's College London, London, UK}
\corremail{jordina.aviles\char`_verdera@kcl.ac.uk}

\fundinginfo{Wellcome Trust fellowship [WT201526/Z/16/Z] and UKRI FLF fellowship [MR/T018119/1], an NIHR Advanced Fellowship to LS [NIHR3016640], Heisenberg funding from the DFG [502024488] and core funding from the Wellcome/EPSRC Centre for Medical Imaging [WT203148/Z/16/Z].\\
\vspace{5mm}
Word Count: 3414}

\runningauthor{Jordina Aviles Verdera et al.}

\begin{document}

\maketitle

\begin{abstract}
\noindent \textbf{Purpose:} To provide real-time quantitative organ-specific information - specifically placental and brain T2* - to allow optimization of the MR examination to the individual patient.

\noindent \textbf{Methods:} A FIRE-based real-time setup segmenting placenta and fetal brain in real-time, performing T2* fitting and analysis and calculation of the centile was implemented. A nn-UNet were trained and tested on 2989 datasets for the fetal brain and a second one trained on 210 datasets for the placenta for automatic segmentation. T2* normal curves were obtained from 106 cases and prospective evaluation was performed on 10 cases between 35 and 39 weeks GA. 

\noindent \textbf{Results:} Quantitative brain and placental T2* maps and centiles were available in all prospective cases within 30 seconds. The robustness of the method was shown with intra-scan repeats (mean difference 1.04$\pm$12.39 ms for fetal brain and -3.15$\pm$8.88 ms for placenta) and direct validation with vendor-processed offline results (mean difference 1.62$\pm$4.33 ms for fetal brain and 0.16$\pm$6.19 ms for placenta).

\noindent \textbf{Discussion and Conclusions:} Real-time available organ-specific quantitative information enables more personalized MR examinations, selection of the most pertinent sequences and thus the promise of reduced recalls and specific insights into tissue properties. The here enabled placental T2*, demonstrated in multiple recent studies to be a biomarker sensitive to a range of pregnancy complications, and fetal brain T2* will be explored in further studies in pregnancies with pre-eclampsia, growth restriction as a way of enabling future MR-guided fetal interventions.

\end{abstract}

\section{Introduction}

\noindent Fetal MRI has an important complementary role to ultrasound (US) in antenatal assessment, monitoring and delivery planning. Specific advantages include high spatial resolution, excellent soft tissue contrast, feasibility in late gestation and importantly a wide range of quantitative and functional contrasts. One of the current limitations of fetal MRI, however, is the lack of real time information obtained which limits timely alterations in patient care. This is particularly pertinent in late gestation scans, where delivery may be warranted depending on findings \cite{Peasley2023-tb}. Although ultrasound is the mainstay of conditions such as fetal growth restriction (FGR), in the research setting MRI has shown potential to provide additional information as, unlike ultrasound, it can provide comprehensive assessment of functional properties of placental tissue, in addition to downstream effects on the fetus. Undetected fetal growth restriction  \cite{Gordijn2016-bg} for example increases the risk of stillbirth four-fold, with late FGR being the most common cause of 3rd trimester loss\cite{Poon2016-tg,Ego2020-ov}. Further tools to identify the fetuses at most risk of adverse outcomes, allowing  timely interventions are likely to have a significant impact on patient care. If identified, timed delivery can prevent late stillbirth \cite{King2022-nr}.\\

\noindent While the detection of reduced anatomical growth during an US scan triggers additional flow assessments in the uterine artery, umbilical vein and superior vena cava among others \cite{Uquillas2017} to evaluate the placental reserve and ability to sustain fetal growth in late gestation, no such real-time change is currently possible in MRI as the analysis is performed after scan.  T2* relaxometry is one such functional contrast which is being increasingly utilised in fetal MRI. It has shown significant potential in placental evaluation as an indirect measure of oxygenation, with mean placental T2* values following a linear decay over gestation \cite{schabel2016functional,Sinding2018,Sorensen2013} in uncomplicated pregnancies with reduced values observed in women with pre-eclampsia (PE) \cite{Ho2020}, and FGR \cite{Sinding2018}. Fetal organ development has also been evaluated with T2* decline shown in most organs, including the lung, brain and thymus  \cite{Sethi2021,Payette2023-gz}. Furthermore, recently low field MRI has been utilised for fetal MRI \cite{aviles2023reliability,ponrartana2023low}. T2* relaxometry is particularly attractive at low field strengths due to the longer T2*, and thus larger dynamic range even in late gestation and cases associated with pathology. Other benefits of low field scanning include: reduced artifacts due to the increased field homogeneity, and improved comfort due to wider bore sizes facilitating assessment of women with high body mass indices (BMI) and those in late gestation. 

\noindent While the acquisition of T2* data is usually fast if performed with multi-echo gradient echo sequences ($<$1 minute), the required organ segmentation and the fitting to obtain quantitative information is typically manual and time consuming.  This also means the opportunity for optimal data acquisition is lost, for example if repeat sequences are required for placental assessment in the presence of a uterine contraction, or fetal data where excessive motion is present.  These factors all limit the potential for functional MRI techniques to provide realtime information to direct patient care.\\

\noindent Recent work has demonstrated the use of artificial intelligence (AI) methods to automatically segment the placental parenchyma \cite{Abulnaga_2023,PIETSCH2021102145,Hall2024} with promising Dice scores. Furthermore, feasibility of online real-time segmentation and quantification has been demonstrated in the pediatric heart \cite{STEEDEN201879} and quantitative online relaxometry utilised in adult cardiac MRI \cite{CHRISTODOULOU2024100997}. In fetal MRI, recent work has enabled real-time changes, concretely updating the field-of-view was shown during the scan for gradient echo scans \cite{Neves2023} and online planning for the fetal brain \cite{hoffmann2021rapid,neves2024fully} and heart \cite{silva2024automaticflowplanningfetal} was demonstrated.\\

\noindent In this work, we present an integrated, automatic, realtime assessment on the scanner of the placenta and fetal brain resulting in segmented, fitted T2* maps and corresponding centiles in relation to control populations. The method was fully deployed online and evaluated prospectively in 10 fetuses over  35 weeks of gestation. Robustness of the real-time pipeline was demonstrated using both intra-scan repeats and comparison with the conventional offline pipeline in all subjects.\\

\section{Methods}

\subsection{Fetal imaging and ethics}
\noindent Women were prospectively recruited to three ethically approved studies: MEERKAT (21/LO/0742 Dulwich REC, 08/12/2021), MIBIRTH (23/LO/0685, London-Harrow REC 14/09/2023) or NANO (19/LO/0736, London-Stanmore REC 20/06/2019). Informed consent was obtained and scans performed on a commercially available 80 cm bore low-field scanner (MAGNETOM Free.Max, Siemens Healthineers, Erlangen, Germany) using a blanket-like 6-element coil (BioMatrix Contour L Coil, Siemens Healthcare, Erlangen, Germany) and a permanent 9-element spine coil. Participants were scanned in a head-first supine position with extra head, back and leg rests as required. If supine position was not tolerated, supported left-lateral position was offered. Continuous heart rate and intermittent blood pressure monitoring were undertaken together with frequent verbal interaction. A break was offered half-way through the examination.\\

\subsection{Training, masking and centile calculation}
\noindent Automatic segmentation of the placenta and fetal brain was achieved using the nn-UNet framework \cite{isensee_nnu-net_2021}. For the fetal brain, a contrast-, field strength-, pathology- and age-agnostic network was trained and tested with a total of 2989 datasets from 1045 participants including  352, 364 and 354 ssTSE datasets, 75, 55 and 46 bSSFP datasets and 282, 314 and 420 T2* mapping datasets acquired at 0.55T, 1.5T and 3T respectively. 120 diffusion volumes from a range of b-values/b-vectors and echo times and 607 T1 mapping volumes were acquired at 0.55T \cite{aviles2023reliability, AvilesVerdera2024}. Ground-truth masks were obtained either manually by a range of fetal MRI experts (3-10 years of experience) or using in-house trained networks. For the placenta, a multi-field strength network was trained with a total of 210 multi-echo gradient echo datasets, 70 at 0.55T, 70 at 1.5T and 70 at 3T \cite{Hall2024}. Ground-truth masks were generated manually using the first echo time (TE) avoiding amniotic fluid and maternal vasculature by a range of fetal MRI experts (3-10 years of experience). In both networks, a split of 80\%/20\% for training and validation was used. \\

\noindent A cohort of 106 healthy control datasets, among these 68 above 32 weeks GA (Cohort \textit{NormalCurves}), were used to generate centiles for both placenta and fetal brain across gestational age (GA). Outcome data was obtained from clinical records to ensure that no antenatal, delivery, postnatal or neonatal complications occurred and that these datasets were suitable as true controls. For the entire cohort, the mean maternal age  was 33.86 years, the mean BMI=27.83kg/m$^2$ and the mean gestational age at scan= 32.7 weeks, for the cohort above 32 weeks, the mean maternal age was 33.55 years, the mean BMI=26.66kg/m$^2$ and the mean gestational age at scan= 36.50 weeks.\\

\begin{figure}[!ht]
    \centering
    \includegraphics[width=0.95\textwidth,clip]{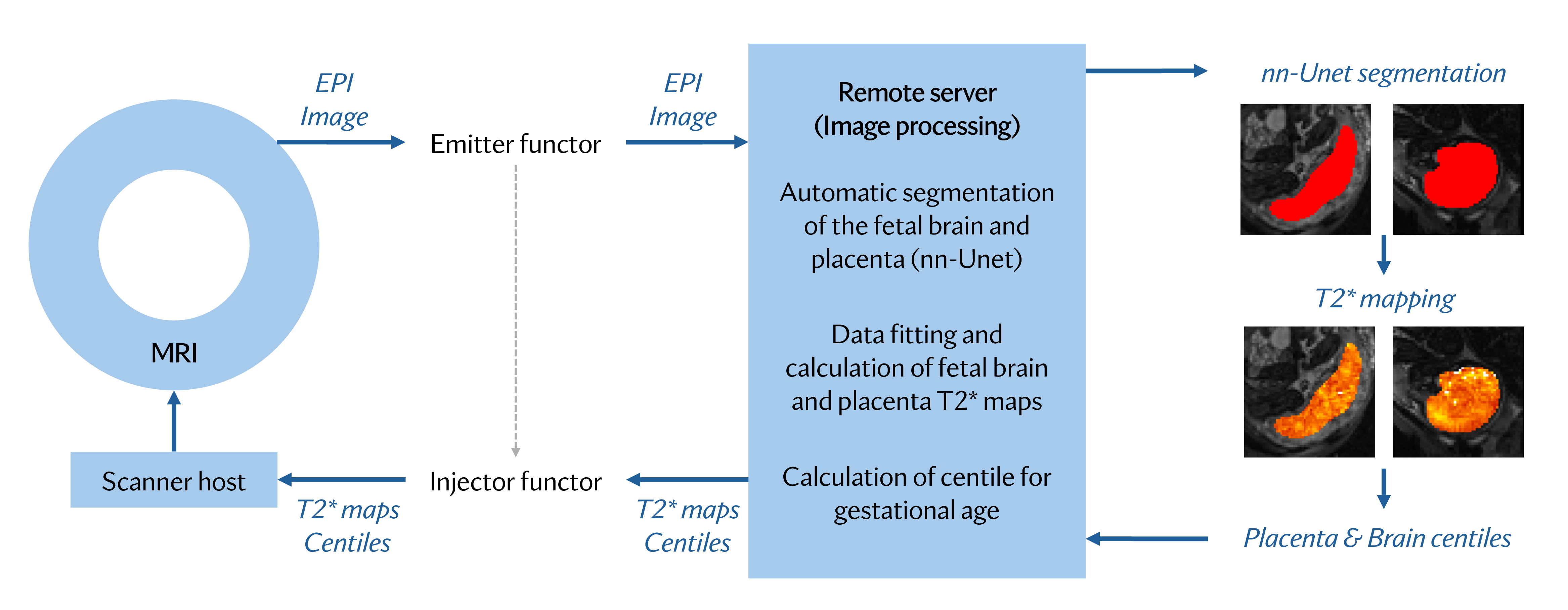}
\caption{Processing pipeline for real-time T2* centiles from the acquisition (left) to the gadgetron processing (mid) to the masks and maps (right).} \label{methodologypipeline}
\end{figure}

\noindent Mean T2* values for both placenta and fetal brain were obtained from the \textit{NormalCurves} cohort using vendor-provided reconstruction, manual segmentation and T2* quantification using least squares in python as previously described \cite{aviles2023reliability}. Linear regression analysis was then performed using the ordinary least-squares module in python to obtain the trend over gestational age. Finally, the 10th, 50th and 90th centiles were calculated for both placenta and fetal brain using quantile regression in python. A classification into low T2* (under the 10th centile), mid T2* (between the 10th and 90th centile) and high T2* (above the 90th centile) was conducted. The prospective entire real-time T2* acquisition and post-processing pipeline is also graphically shown in Figure \ref{methodologypipeline}.\\

\subsection{Real-time cohort and validation}
\noindent The complete described real-time implementation was tested online in a total of 10 participants (Cohort \textit{RealTime}) above 35 weeks GA to focus on the clinical question of late onset FGR. To ensure robustness and accuracy of the presented pipeline, two different validation methods were applied for the \textit{RealTime} cohort. First, a total of three acquisitions were obtained and processed real-time for each dataset to ensure robustness to different time points and maternal or fetal movement. To evaluate accuracy, automatic segmentation, T2* mapping and centiles calculation was also performed offline in the scanner reconstructed data in parallel to the real-time pipeline.\\

\section{Results}
\begin{figure}[!ht]
    \centering
    \includegraphics[width=0.95\textwidth,clip]{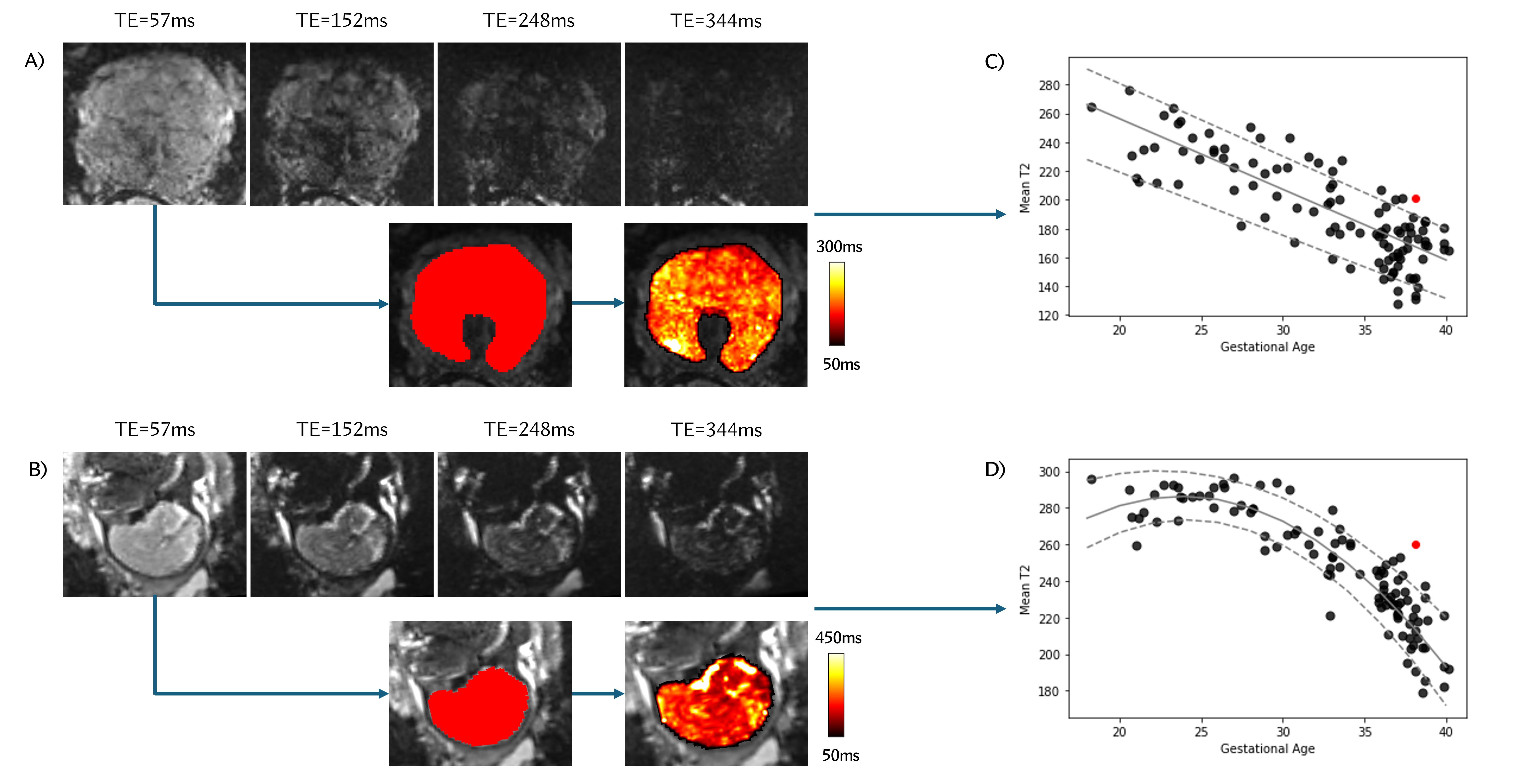}
\caption{Complete overview of all information available in real-time for one selected case for (A) the placenta and (B) the brain. The images at all four echo times are shown, the organ-specific masks in red, the T2* map and the quantitative mean T2* value plotted against the normal cohort.} \label{results_beautiful}
\end{figure}

\begin{figure}[!ht]
    \centering
    \includegraphics[width=0.95\textwidth,clip]{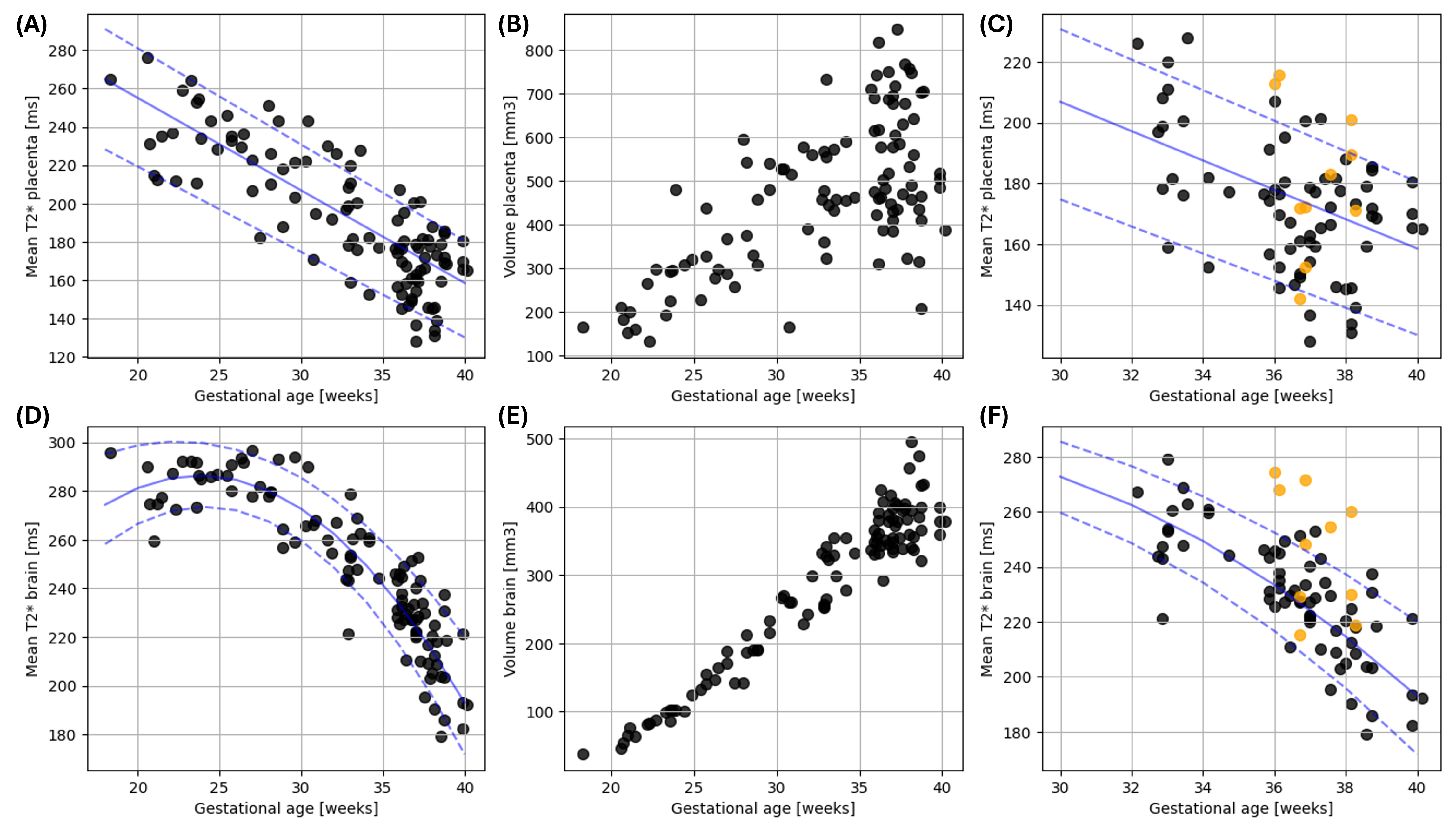}
\caption{Quantitative obtained T2* results from both cohorts (from the normal cohort only the 0.55T data was taken) plotted against gestational age. In black the \textit{NormalCurves} cohort for the placental mean T2* entire gestational age range (A,C), for the weeks from 32 to 42 weeks (B,D), in blue the calculated 10th, 50th and 90th centile lines and and in orange the data from the \textit{Real-Time} cohort.} \label{results_quant}
\end{figure}

\noindent The described pipeline was successfully implemented into the scanner online processes and tested prospectively. Data was acquired in a total of 10 cases in real-time (Mean GA= 37.13 weeks [36,38.28], mean BMI at scan= 29.35 kg/m2 [21.17,37.51] and mean maternal age at scan=33.6 years [30,36]). Figure \ref{results_beautiful} visualizes a complete example with data acquired at multiple echo times, the automatically obtained masks, the T2* fits in brain and placenta and finally the resulting quantification against normal curve values for one of the included participants. Fig \ref{results_quant} displays the T2* and volume results for the \textit{RealTime} cohort plotted against GA and overlaid on the normative curves at low-field obtained with the \textit{NormalCurve} cohort and a closer look to the late gestation range.\\

\begin{figure}[!ht]
    \centering
    \includegraphics[width=0.99\textwidth,clip]{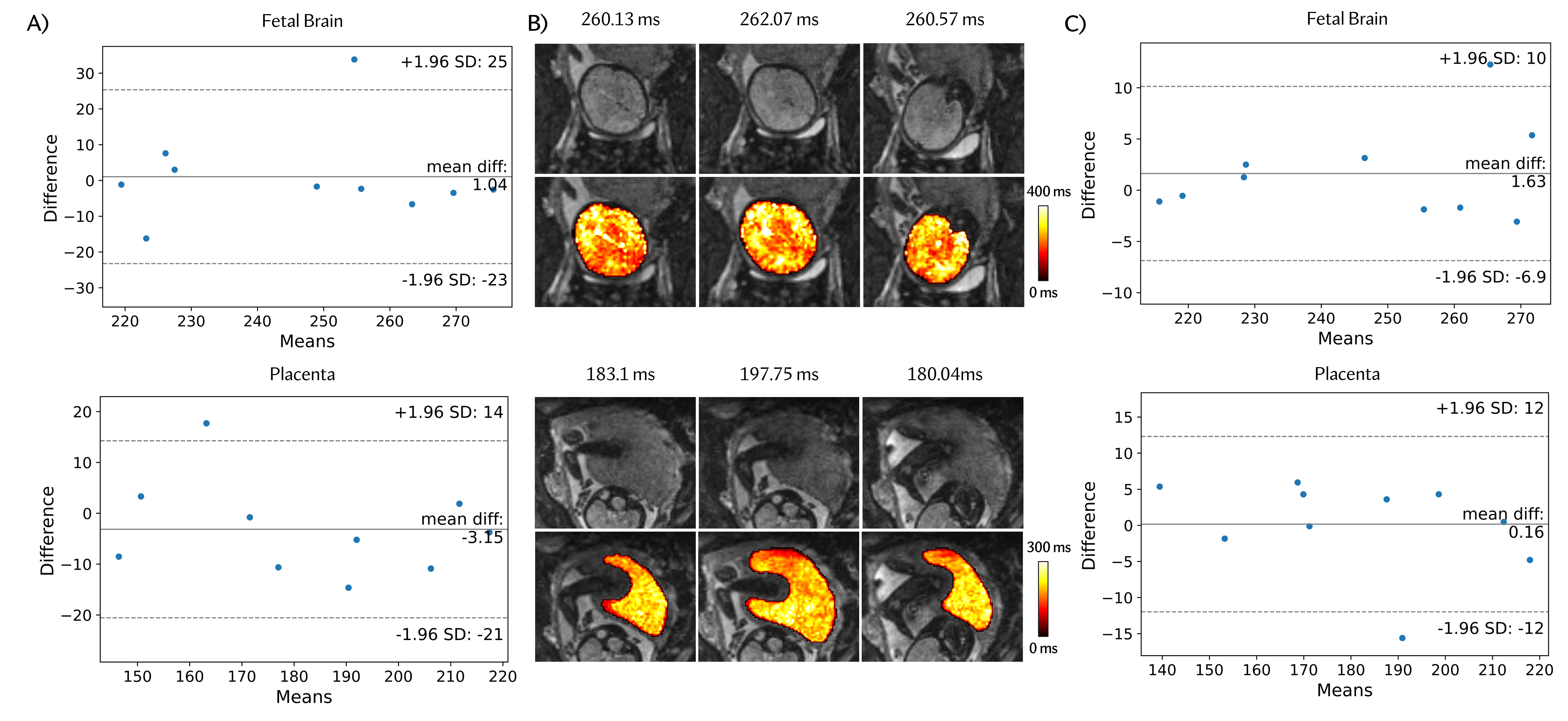}
\caption{Quantitative results of the robustness study. A) Bland-Altman plot between first and second real-time T2* calculated maps, B) Coronal view of first-echo time with its corresponding maps for all real-time acquisitions in an example at 37 weeks, C) Bland-Altman plot between first real-time acquisition and subsequent offline vendor reconstructed data.} \label{results_robust}
\end{figure}

\noindent The results from the robustness study are displayed in Fig. \ref{results_robust} with A) a Bland-Altman plot with a mean difference of 1.04+-12.39ms for the fetal brain and -3.15+-8.88ms for the placenta between the first and second acquisition together with B) first-echo coronal view of the fetal brain and placenta with its corresponding maps for all the acquisitions in an example at 37 weeks and C) a Bland-Altman plot with a mean difference in T2* of 1.63+-4.33ms for the fetal brain and 0.16ms+-6.19ms for the placenta between the first real-time acquisition and the offline vendor reconstructed data, acquired sequentially. Robustness analysis between the other acquisitions show similar results to the ones presented in Fig. \ref{results_robust} with mean differences in T2* of -3.17+-13.61ms for the fetal brain and -4.74ms+-7.33ms for the placenta between first and third acquisition and -5.07+-7.23ms for the fetal brain and -2.45+-8.34ms for the placenta between the second and third acquisitions. Placental centiles remained unchanged between repetitions in all but four cases, three due to contractions and the other due to sub-optimal segmentation in high BMI. For fetal brain, centiles remained unchanged between repetitions in all but three cases, all related to artifacts due to fetal head motion. Centiles between the two subsequent acquisitions used to compare online and offline pipelines remained unchanged for both organs in all cases. 

\section{Discussion}
\noindent A fully automatic real-time online pipeline was introduced for online organ-specific T2* quantification from a short ($<$30sec) whole-uterus multi-echo gradient echo scan. Placental and fetal brain segmentation with corresponding quantitative T2* maps and centiles were obtained in under a minute. The pipeline worked reliably in all tested cases, produced accurate segmentations in all cases for the fetal brain and all but one case for the placenta (where maternal fat had been included in the segmentation). Quantitative T2* values for both brain and placenta were in-line with previously published results on the same field strength \cite{aviles2023reliability} and followed trajectories as shown in previous publications at higher field strengths \cite{Sorensen2013,schabel2016functional}.\\

\noindent The presented normative curves show a reduction in mean T2* reduction for both organs over gestation which is in accordance with previous published work \cite{sorensen_2020}. The higher absolute values observed in our study are likely due to different field strengths used. The fetal brain volumes show the expected significant positive correlation with gestational age while the placenta had a general trend of increasing volume with increasing gestational age but with higher variability particularly at late gestations. Differences found in T2* between the present study our previous study published at 0.55T \cite{AvilesVerdera2024} are likely attributable to the larger sample size, especially towards late gestational age. While mean fetal brain T2* follows a quite consistent trend throughout pregnancy, mean placental T2* values at late gestation (above 32 weeks) become more scattered. This could be related to the increased frequency of contractions at this GA or related to approaching a time frame where the variability of delivery, with term ranging from 27 to 41 weeks, plays an increased role. However, there is limited availability with normal T2* values for the placenta within a large sample of data above 32 weeks that reproduce this variability and present an accurate explanation.\\ 

\noindent The robustness results show the consistency of the real-time pipeline within three different multi-echo datasets at different time points within the whole acquisition. Two cases had a difference higher than 10ms difference for placental T2*, with detailed investigation revealing that the first one captured contractions in two of the acquisitions (see Fig. \ref{results_robust} B) causing a reduction in T2* and the second case can be explained with increased geometric distortion in one of the acquisitions. The two cases with a difference higher than 10ms for the fetal brain were visually related to high motion patterns. The other cases showed minimal differences for both placenta and fetal brain within acquisitions with consistent centiles across them. The robustness shown between the presented pipeline and the offline vendor reconstructed pipeline proves the accuracy of the real-time automatic assessment of multi-echo data, with only one case with a difference higher than 5ms for the placenta due to a contraction. Centiles changed between repetitions in four cases for placenta and three cases for fetal brain. However, the robustness analysis showed no significant differences between repetitions meaning small variations around the limit of the centile might make the centile change. Furthermore, changes were explained by contractile activity happening in the case of the placenta and fetal head motion in fetal brain, suggesting identifying these changes can trigger new acquisitions

\paragraph{Strengths and limitations}
\noindent The availability of the masks, T2* maps and centiles in real-time while the scan is still ongoing is a key strength of this study, as it allows the ddetection of  abnormal values not only due to pathologies but also secondary to contractions or fetal motion during the scan as demonstrated in the presented cases. This enables an immediate reaction, for example triggering the acquisition of extra sequences that provide additional information or the re-acquisition of the sequence in case of contraction or motion corrupted images. The accurate immediate detection of low T2* informs the clinicians immediately, essential especially at late gestational age, for the detection and treatment of late onset FGR and PE - two of the most common causes of stillbirth \cite{peasley_2023, king_2022, ashoor_2022} allowing timely intervention. The availability of all this information in less than a minute opens up the possibility for future clinical applications where real-time feedback is key. A promising application in the future could be twin-to-twin transfusion syndrome surgery. While previous work using fetal MRI has been undertaken in the context of planning ablation surgeries, no real-time information was available \cite{luks_2001, torrents-barrena_2019}. The ability to perform this under image guidance with real-time quantitative values might help to accurately assess the effects of the ablation in-situ. \\

\noindent However, this study also has limitations. The accuracy of placental segmentation in cases with higher BMI requires further improvement, as currently additional tissue elements in subcutaneous fat have been wrongly identified as placental tissue. Further work will include integrating the online T2* maps  in a separate graphical display to further facilitate the real-time reaction of the radiographer or the scanner. Furthermore, the \textit{RealTime} cohort sample size could be enhanced by  FGR or PE cases. 

\paragraph{Implications and next steps}
\noindent Next steps include the better integration of the achieved real-time information into the scanner console to improve  the clinical workflow even further. This would also enable the triggering of additional sequences in  the presence of pathology, contraction or motion. In addition, an  increase in sample size with a higher variability will allow a more accurate assessment of pathological cohorts such as pregnancies with PE or late FGR. A retrospective analysis of previously acquired offline T2* available datasets together with delivery outcomes will be conducted and will help tailoring the pipeline in addition towards early abnormality detection. Moreover, including additional structural and quantitative measures such as placental location or texture may provide  a more in depth assessment  of placental function in real-time. The method was demonstrated here for 0.55T, the applicability is however wider and will be tested in the next steps on higher field MR scanners. Finally, the achieved real-time setup paves the way for further contrasts or organs to be quantified in real-time.

\section{Conclusion}
\noindent The availability of quantitative organ-specific T2* maps together with centiles based on a well characterized cohort  provides  a more interactive fetal MRI examination, tailoring the scan for the individual, the ability to  react to image findings immediately and thus to provide more timely information for antenatal care. This approach is ideal for wider usage and could enable  disease specific biomarkers  in the future.

\section*{Acknowledgements}
The authors thank all pregnant women and their families for taking part in this study. The authors thank the research midwives and obstetric fellows for their invaluable efforts in recruiting and looking after the women in this study as well as perinatal radiographers for their involvement in the acquisition of these datasets. This work was supported by a Wellcome Trust Collaboration in Science grant [WT201526/Z/16/Z], a UKRI FL fellowship [MR/T018119/1] and DFG Heisenberg funding [502024488] to JH, an NIHR Advanced Fellowship to LS [NIHR3016640] and by core funding from the Wellcome/EPSRC Centre for Medical Engineering [WT203148/Z/16/Z]. The views presented in this study represent these of the authors and not of Guy's and St Thomas' NHS Foundation Trust.



\newpage
\clearpage

\paragraph{Figures and Tables}

\end{document}